\documentstyle[11pt]{article}
\input FEYNMAN
\title{\vspace{-1.in} \hfill {\small\rm TUM-HEP-348/99} \\~\\~\\
Strategy towards Mirror-fermion Signatures}
\author{ George Triantaphyllou\thanks{e-mail:georg$@$ph.tum.de}
$\;$\\~  \\{\it Institut f\"ur Theoretische Physik, Technische 
Universit\"at M\"unchen}\\
{\it James-Franck-Strasse, D-85748 Garching, GERMANY }}
\begin{document}
\setlength{\baselineskip}{24pt}
\maketitle
\thispagestyle{empty}
\begin{abstract}
The existence of mirror fermions interacting strongly under a new
gauge group and having masses near the
electroweak scale has been recently proposed as a viable  
alternative to the standard-model Higgs mechanism. The main purpose 
of this work  is to investigate which specific experimental 
signals are needed to clearly differentiate the mirror-fermion model
from other new-physics models. 
In particular, the case is made for a future large
 lepton collider with c.o.m. energies 
of roughly  4 TeV or higher. 
\vspace{2.in}
\end{abstract}
\vfill
\setcounter{page}{0}
\pagebreak
\renewcommand{\thefootnote}{\alph{footnote}}
\markright{} 

\section {Introduction}
Most of the current high-energy experimental data are in good agreement with
the theoretical predictions of the
standard model of elementary particle physics. This model predicts
however the existence of a fundamental 
scalar field, the Higgs particle, having  a mass on the
order of the electroweak scale, which has yet to be discovered. The instability 
of the mass of an elementary scalar field against quantum corrections 
nevertheless
has prompted various speculations regarding the true inner structure of 
the Higgs sector along the years, 
leading to the study of new physics beyond the standard model. 

It has been recently argued \cite{george1} that the standard Higgs
mechanism can be seen as an effective low-energy description of a new 
heavy sector
consisting of mirror fermions carrying quantum numbers under a 
gauged generation group becoming strong at around 1 TeV. Non-zero
vacuum-expectation values of mirror-fermion bilinear operators
can then break the
electroweak symmetry dynamically at the expected energy scale. 

The subsequent breaking of the
mirror-generation group allows the formation of composite fermion 
operators which are invariant under the
electroweak 
gauge symmetry. These operators feed masses to the standard-model fermions by 
mixing them with their mirror partners in a way that can suppress dangerous 
flavor-changing neutral currents. The resulting masses and mixings for
quarks and leptons can easily accommodate current experimental results.

This model \footnote{people tending to introduce new terms
could name it  ``mirrorcolor", and the new
fermions ``katoptrons" from the greek word meaning ``mirror"} 
presents from the theoretical side several advantages. 
The natural solution offered to the hierarchy problem is accompanied by an
appealing gauge-group unification including the mirror-generation group. 
It is quite important to stress
here that the naturalness problem, far from being
just a matter of stability of the electroweak scale with 
respect to the Planck scale against radiative corrections, 
has to do mainly with the fact that the
electroweak scale is roughly 14 orders of magnitude, and not some other
arbitrary number, smaller than the gauge-coupling unification
scale. The mirror model, unlike most other currently popular  
new physics approaches, addresses this issue in a very satisfactory manner,
without  resorting to the anthropic principle for instance. 
However, a deeper understanding of the mechanism that is responsible
for the eventual breaking of the strong mirror-generation group is still
needed.

The gauge-coupling unification in this model is found to be
not only consistent with proton-decay bounds,
but can also lead to a precise explanation of the order of magnitude of the
QCD and electroweak scales without any need for fine tuning of 
parameters \cite{george2}. 
Moreover, the introduction of mirror partners to
the known fermions restores in a certain sense the chiral 
symmetry missing in the standard model, and in addition
could constitute a solution to the 
strong CP problem \cite{Barr}. 

From the experimental side, electroweak precision tests  could already be 
providing indirect signals for physics beyond the standard model in
this direction. Of 
particular importance are
here the almost 3$\;\sigma$ deviations of the right-handed weak
coupling of the bottom quark \cite{Field} and the values of the $S$ and $T$
parameters, which, even though still consistent with zero, can take 
non-negligible negative values. Contrary to most other new-physics models,
these effects can be explained within the mirror-model framework. 
Still missing nevertheless is a
deeper theoretical understanding of why certain fermion composite 
operators take the particular 
values which make the theory consistent with these effects. 

Having described  the main features of the mirror model, 
an effort should be
made to see how it can be tested phenomenologically. Of special interest
is of course the search for signals which clearly differentiate it from 
alternative theories, and to investigate what types of high-energy
facilities would be required for such an endeavor. Far from being
an exhaustive or thorough study, this work tries to sketch roughly
the logic which,
taking into account what is feasible presently and in the
years to come experimentally, can lead to a definite verification
or falsification of the model. We hope to return to the
specific processes described here in order to study them 
in more detail in the future. 

\section{Indirect tests}

\subsection{Generic features}  

The existence of new heavy particles can in principle influence
the couplings and masses of the standard-model particles
via higher-dimensional operators. The corresponding effects should be usually
detectable at lower energies in anomalous decays of the known particles, or
in physical quantities taking values deviating substantially from their
standard-model predictions, without having to 
probe the heavier
new-physics sector directly. The disadvantage of this program
is however that various distinct
theories can frequently predict similar effects. 
The information drawn by such analyses is useful therefore mainly to the
extent of constraining the respective
model parameters, and possibly ruling out whole classes of theories, 
but  not proving unambiguously that a particular model is correct.

As a typical example, one can quote proton decay. It was recently shown
\cite{george2} that, if mirror-fermion models are to be compatible with 
gauge-coupling unification, they should predict proton-decay 
like $p \longrightarrow e^{+}\pi^{0}$
with rates close to current experimental bounds, as given from the 
Super-Kamiokande experiment for instance. 
In particular, for the favored Pati-Salam gauge symmetry-breaking sequence 
scenario 
\begin{equation}
SU(4)_{PS}\times SU(2)_{R} \longrightarrow SU(3)_{C} \times U(1)_{Y}
\end{equation}

\noindent 
one calculates the following approximate
values for the unification scale and common gauge
coupling: 
\begin{equation}
\Lambda_{GUT} \approx 10^{15.5} \;
{\rm GeV},\;\;\;  \alpha_{GUT} \approx 0.036.  
\end{equation}

 The value of the unification coupling
quoted above is quite close to the present experimental limit
of $\alpha_{GUT}~^{<}_{\sim} 0.074$ for the same unification 
scale \cite{george2}. 
Even though detection of proton decay would be of utmost
importance for the generic verification of 
unification schemes, it would not be able by itself to
indicate which particular type of unification is favored, since there are
several other frameworks (like supersymmetric unification for 
example) predicting similar
proton lifetimes. 

Another example is provided by the recently measured
anomalous bottom-quark right-handed weak
 coupling $\delta g^{b}_{R}$
\cite{Field}, or the negatively-centered 
values of the electroweak precision parameters $S$ and $T$, which in
their turn could be 
related to an anomalous top-quark right-handed weak  coupling
\cite{george1}. Unfortunately, this coupling is still directly
unaccessible in current experiments due to the heaviness of the top quark.
Even though the mirror-fermion model  is one of the 
very few examples of theories consistent with such effects, these phenomena
can only serve as an indication and not a definite proof, since in any case the
corresponding deviations of  $\delta g^{b}_{R}, S$ and $T$
are smaller than $3 \;\sigma$. 

With regard to the $S$ parameter in particular, 
it should be noted that the top-quark anomalous coupling 
does not have to be as large as the one quoted in \cite{george1} in
order to cancel the large ``oblique" corrections coming from the
numerous new electroweak doublets introduced in the mirror model. Since the
strong mirror-generation group is  broken, one can speculate that, even though
it forms condensates
and generates dynamical mirror-fermion masses,  no 
vector resonances are formed, in which case the $S$ parameter
is given by 
\begin{equation}
S^{0} \approx \frac{N}{6 \pi}, 
\end{equation}

\noindent with $N$ the number of new heavy electroweak
doublets. This is half as large as the estimate based on QCD-like
dynamics.

Such a scenario has been studied in \cite{Hil}, giving
a result consistent with the one of Ref.\cite{yale} for the case of roughly
momentum-independent fermion self-energies. 
The reason for the appearance of non-QCD dynamics here is
however not a ``walking" gauge coupling but a broken gauge group.
In the mirror-model case, the number of new weak doublets introduced is
$N=12$, so one estimates $S^{0}\approx 0.64$. 
The presence of Majorana mirror neutrinos can make $S$ even
smaller \cite{george2}, since the leptonic contribution can be
as small as $S^{0}_{l}\approx -0.24$ instead of
$S_{l}^{0} \approx 0.16$ for the case of
mirror neutrinos and charged leptons which are 
degenerate in mass. This ``best-case"
scenario leads to a total
``oblique" contribution to the $S$-parameter given by $S^{0}=
S^{0}_{q}+S^{0}_{l}\approx 0.24$, where $S^{0}_{q}$ stands for the 
mirror-quark oblique contribution to the $S$ parameter.  

In the mirror model, contributions of vertex corrections to the
$S$-parameter coming from effective four-fermion operators
can also be potentially important. In order
to be within 1 $\sigma$ from  the experimental limit on $S$ quoted in 
\cite{george1}, one needs
the vertex corrections to  give a contribution to the $S$ parameter 
on the order of 
\begin{equation}
S^{t,b} ~^{<}_{\sim} - 0.09, 
\end{equation}

\noindent which is about one order of magnitude smaller
than the value used in \cite{george1}, and in absolute value
not unreasonably large.   In fact, if one accepts
the current experimental central value for $\delta g^{b}_{R}$, 
it could be produced by an anomalous top-quark coupling of
about $\delta g^{t}_{R} \approx  - 0.03$, which is roughly equal 
in absolute value to $\delta g^{b}_{R}$. 
A small top-quark anomalous coupling can therefore easily accommodate the
present experimental data. 

The reduction of the needed magnitude for  
$\delta g^{t}_{R}$  leads in addition to the elimination
of excessive fine tuning needed to keep the $T$ parameter 
small, since in \cite{george1} this
parameter is mainly affected by vertex corrections induced by
$\delta g^{t}_{R}$. The isospin breaking introduced artificially 
within the mirror doublets in that reference in thus rendered obsolete in
this scenario. 

Oblique corrections to the $T$ parameter
in these models are generally expected to  be small since
the mirror-fermion masses are dynamically generated
and roughly isospin symmetric in analogy to the constituent quark masses in
non-perturbative QCD, and  the difference of the standard-model 
top- and bottom-quark masses is taken to be fed down in a gauge-invariant way. 
In the mirror-lepton sector things are more complicated however, since
the see-saw mechanism responsible for the mass splitting between
the charged mirror leptons and the mirror neutrinos which generates 
the negative value of $S^{0}_{l}$ generates also positive contributions
to the $T$ parameter. These would still have to be roughly canceled  by
the $\delta g_{R}^{t}$-contributions
to accommodate current experimental limits.

Particularly useful in this respect will be direct measurements of 
the anomalous coupling
$\delta g^{t}_{R}$ via the top-quark forward-backward asymmetry in 
the Next Linear Collider (NLC). 
Following the analysis of Ref. \cite{Chivu} regarding
experiments at the NLC, the top quark neutral-current
coupling can be constrained there at the $10\%$ level, which 
should be enough
to discover mirror-fermion mixing effects.  This coupling could also
be constrained in the planned muon collider. 

Another important quantity which can  deviate substantially 
in the mirror model from its value predicted by the standard-model is  
the CKM matrix element $|V_{tb}|$. 
In the numerical example presented in \cite{george1}, it was found that 
\begin{equation}
|V_{tb}| \approx 0.95. 
\end{equation} 

\noindent This quantity can be made closer to unity for 
heavier mirror top quarks, but it cannot exceed by a lot 
the value quoted above if one wants to reproduce the weak scale
correctly.  Deviations from the standard-model
prediction of $|V^{SM}_{tb}| \sim 1$  would support
the existence of at least one new fermion generation mixing with the
known fermions in order to 
guarantee the unitarity of the generalized mixing matrix. 

The value of $|V_{tb}|$ could be tested
via virtual W-boson and top-quark decays at 
the Tevatron III and the NLC. 
At these high-energy
facilities, a limit of $|V_{tb}|>0.97$ could be obtained if
the standard-model value is correct \cite{Chivu}, providing a 
good testing ground for the mirror model.  

A deviation of $|V_{tb}|$ from its standard-model value
would provide a hint for a mass-generating mechanism for the
ordinary fermions via feed-down (a generalized ``see-saw")
from the new sector.   In conjunction with
anomalous heavy-quark weak couplings however, it could further be an
indication of mixing with at least one  new fermion generation having 
different weak-charge assignments
from the standard-model generations, and  would thus
lend support to the mirror-fermion framework proposed in \cite{george1}.
The new-generation fermions could of course
be {\it a priori} weak singlets and not directly involved in the
electroweak-symmetry breaking, so it is  
still required to see whether they decay weakly. 

\subsection{Flavor-changing neutral currents}  

Theories introducing mirror fermions usually predict the existence of 
flavor-changing neutral currents (FCNC) \cite{Roos}. These are expected to
be particularly important for processes involving  heavier quarks, 
since in the mirror model
the masses of the known fermions are generated by their mixing  
with their mirror partners. 

The mixing of the two lighter generations with their 
mirror counterparts is quite small, as can be seen
in the generalized CKM
matrices for quarks \cite{george1} and leptons \cite{george2}.
Therefore, $K^{0}-\bar{K^{0}}$ mixing, as well as decays like  
$K \longrightarrow \pi \nu \bar{\nu}$, $K \longrightarrow e \mu$, 
$\mu \longrightarrow e \gamma$ and  $\mu \longrightarrow e \nu \bar{\nu}$
can be  made to
agree with present experimental limits without much effort. 
The same can be said about the CP-violation
parameters $\epsilon$ and $\epsilon^{\prime}$ in the kaon system, 
even though it is still
conceivable that the recently reported deviation of 
$\epsilon^{\prime}/\epsilon$ from the standard-model prediction coming
from the KTeV experiment is related to such type of physics. 
Effects coming from the new sector can be usually
suppressed by raising the mirror-fermion masses and thus decreasing somewhat
the corresponding mixing. 

This could not be said for the FCNC   
processes involving both third-generation quarks $t$ and $b$
like $ B \longrightarrow X_{s}
\gamma$ or $X_{s}l {\bar l}$, $B_{s} \longrightarrow l^{+}l^{-}$, or for the
$B^{0}_{s} - {\bar B^{0}_{s}}$ mixing which 
could be measured with some accuracy
at the B-factories and the LHC-B. The same goes for the top-quark decays
$t \longrightarrow c \gamma, c Z^{0}$ which could be seen
at the Tevatron III, and the off-shell $Z^{0}$-boson decay  
$Z^{0\;*} \longrightarrow t c$ at the NLC. Since the mass eigenstate
corresponding to the top quark has a non-negligible mirror-top-quark
admixture, one could expect potentially 
measurable effects coming from  these processes. 
On the other hand, processes like 
 decays induced by mirror-lepton 
mixing with standard-model leptons, 
like $Z^{0\;*} \longrightarrow \nu \nu^{M}$ 
\footnote{Following the convention of
\cite{george1}, from now on the mirror fermions 
 are denoted by the symbol of their standard-model partners
 with a superscript ``M".} 
are expected to be highly suppressed due to the small mixing of 
the leptons with their mirror partners 
and not easily detectable at experiments in the NLC for instance. 

 An example on the process $B \longrightarrow X_{s} \gamma$ induced by
a W-boson loop is presented in 
the following, since there exist currently precise results from the
CLEO collaboration on the relevant 
branching ratio, and since experimental data for the other processes are not
yet available or not precise enough to lead to useful model constraints. 
Extending the formalism of  Ref.\cite{Cho} to encompass mirror quarks, 
the inclusive rate for $B$-meson 
decay modeled upon the quark process $b \longrightarrow s \gamma$ is
given by
\begin{equation}
\Gamma(B \longrightarrow X_{s} \gamma) = 
\frac{8}{144\pi^{2}}G^{2}_{F}m_{b}^{5}
\alpha|V_{tb}V^{*}_{ts}(C(m_{t})+C^{\prime})
+\sum_{i}V_{ib}V^{*}_{is}C(m_{i})|^{2}, 
\end{equation}

\noindent where $G_{F}$ and $\alpha$ 
are the Fermi and fine-structure constants
respectively, 
$m_{b}$ the mass of the bottom quark,   
and $V_{ij}$ the generalized 
CKM matrix elements. The subscript $i$ is running over 
the  mirror quarks in the $W$ loop. Their  
 masses $m_{i}$ always satisfy the relation $m^{2}_{i} \gg m_{W}^{2}$. 

In this formalism, 
the functions $C(m_{i})$ and $C^{\prime}$ are given at leading order by  
\begin{eqnarray} 
C(m_{i}) & = & \eta^{16/23}
\left(\frac{-1 - \frac{5\delta}{8} 
+\frac{7\delta^{2}}{8}}{(1-\delta)^{3}} - \frac{(\frac{9\delta}{4}-
\frac{3\delta^{2}}{2})\ln{\delta}}  
{(1-\delta)^{4}}\right) \nonumber \\
& & + 8(\eta^{16/23}-\eta^{14/23})
\left(\frac{\frac{1}{8} - \frac{5\delta}{8} 
-\frac{\delta^{2}}{4}}{(1-\delta)^{3}} - \frac{3\delta^{2}\ln{\delta}}
{4(1-\delta)^{4}}\right) \nonumber \\
C^{\prime} & = &  3\sum_{j=1}^{8}h_{j}\eta^{p_{j}}, 
\end{eqnarray}

\noindent where $\delta = (\frac{m_{W}}{m_{i}})^{2}$, 
$\eta=\frac{\alpha_{s}(m_{W})}{\alpha_{s}(m_{b})}$ with $\alpha_{s}$ the
QCD coupling, $m_{W}$ the $W$-boson mass and the constants
$h_{j},p_{j}$ can be found in \cite{Cho}. Recent analyses have 
proceeded to more precise next-to-leading-order (NLO)
results \cite{Buras}, but they have an accuracy much higher 
than  the one  needed for the purposes of this study
 given the uncertainties of the mirror-fermion masses and mixing angles. 

It is further assumed that the generalized
CKM matrix elements encode all the information
needed to study this decay in the context of the mirror model. 
In particular, the effects of the mirror-top-bottom
matrix element $|V_{t^{M}b}|$ 
 would correspond in a certain sense to the effects stemming from  the
anomalous coupling 
$f^{tb}_{R}$ in the analysis of \cite{Fuji}. In the standard-model,
the $b \longrightarrow s \gamma$
process at the weak scale is dominated  by the diagram with a top quark inside
the $W$-boson loop due to the large CKM matrix element $|V_{tb}|$ and the 
large top-quark mass. 

In the mirror
model however, the mixing-matrix
element $|V_{t^{M}b}|$ is also non-negligible. Even 
though  the quantity $|V_{tb}V_{ts}^{*}|$ is still quite larger than
$|V_{t^{M}b}V_{t^{M}s}^{*}|$, the diagram with a mirror top quark
$t^{M}$ inside the loop is enhanced due to the first term
involving a large logarithm appearing in 
the quantity $C(m_{t^{M}})$, since the mirror top quark
is quite heavy. In the numerical example in \cite{george1}
for instance, it has been taken to have
a mass equal to $m_{t^{M}} \approx 810$ GeV. 

Note that there is in principle
also a loop diagram involving a $Z^{0}$ boson 
contributing to this process, since there are
flavor-changing neutral currents induced at tree level due to the
mixing of the fermions with their mirror partners. The relevant 
couplings here involve the particles 
$(Z,b,b^{M})$ and $(Z,s,b^{M})$, but their are
suppressed due to the small mixing terms  in the  corresponding 
mass matrix, so the contribution of this diagram is expected to be
negligible. Possible additional 
contributions coming from scalar bound states \footnote{Now and in 
the following we will abusively refer to zero-spin fields generically 
as scalars even if they are strictly speaking pseudoscalars.} of
mirror fermions are also neglected by taking them to be heavy, since
the ones with the larger couplings to the fermions  
of relevance here involve a heavy mirror top quark.

Of interest in the following is the ratio $R$ of the mirror-model to the
standard-model decay-rate prediction, which by virtue of Eq.6 and the 
preceding discussion is equal to  
\begin{equation}
R = \frac{\Gamma_{M}(B \longrightarrow X_{s} \gamma)}
{\Gamma_{SM}(B \longrightarrow X_{s} \gamma)} \approx
\frac{|V_{tb}V_{ts}^{*}(C(m_{t})+C^{\prime})
+V_{t^{M}b}V_{t^{M}s}^{*}C(m_{t^{M}})|^{2}}
{|V^{SM}_{tb}V_{ts}^{SM*}|^{2}(C(m_{t})+C^{\prime})^{2}}, 
\end{equation}

\noindent where the superscript $``SM"$ stands for the 
CKM matrix elements expected from  the standard model, and it is
assumed that the decay is dominated by diagrams with a top and a
mirror-top quark inside the $W$-boson loop. 

The relative interference phase $\omega$  
between the products of the relevant generalized CKM matrix
elements in Eq. 8 is then given by
\begin{equation}
\omega = {\rm arg}(V_{tb}V_{ts}^{*}/V_{t^{M}b}V_{t^{M}s}^{*})
 = \arccos{\left(\frac{{\tilde R} -1 -\rho^{2}}{2\rho}\right)} 
\end{equation}

\noindent
with 
\begin{equation}
{\tilde R} = R\left|\frac{V^{SM}_{tb}V_{ts}^{SM*}}{
V_{tb}V_{ts}^{*}}\right|^{2}, \;\;\; 
\rho=\frac{|V_{t^{M}b}V_{t^{M}s}^{*}|C(m_{t^{M}})} 
{|V_{tb}V_{ts}^{*}|(C(m_{t})+C^{\prime})}.  
\end{equation}

The experimental collaboration CLEO recently reported the value \cite{cleo} 
\begin{equation} B_{{\rm exp}}(B \longrightarrow X_{s} \gamma) = 
(3.15 \pm 0.54)\times 10^{-4} 
\end{equation}

\noindent 
for the branching ratio corresponding to this decay,  where the error
includes both systematic and statistical contributions.   
Theoretically, within the framework of 
the NLO calculation, it is expected that \cite{Buras} 
\begin{equation}
B_{{\rm th}}(B \longrightarrow X_{s} \gamma) = (3.28 \pm 0.30)\times 10^{-4} . 
\end{equation} 

In the process of testing  
the mirror model, the quantity $R$ could  be also seen as
the ratio of the experimental result to the theoretical prediction. 
Therefore,  this ratio  can be given by the relation  
\begin{equation}
R=\frac{B_{{\rm exp}}}{B_{{\rm th}}} = 0.96 \pm 0.26 . 
\end{equation}

\noindent 
This result  can be  readily translated into a
bounded phase $\omega$. 
To provide an indicative example of how one could constrain the  mirror 
model, we use specific numerical values for the relevant quantities, 
ignoring the uncertainties stemming from the mixing-matrix elements and
the fermion masses. 

The mixing-matrix values in the numerical example of \cite{george1} are
given by  
\begin{eqnarray}
|V_{tb}^{SM}|&\approx& 1 \nonumber \\ 
|V_{tb}|&=&0.95 \nonumber \\ 
|V_{ts}| &\approx& |V_{ts}^{SM}| \approx 0.038 \nonumber \\    
|V_{t^{M}b}|&=&0.32 \nonumber \\
|V_{t^{M}s}|&=&0.016 . 
\end{eqnarray}

\noindent
Moreover, by virtue of Eq.7 one computes the values  
\begin{equation}
C(m_{t^{M}}) = - 0.66,\;\;\; C(m_{t}) = - 0.42, \;\;\; 
C^{\prime} = - 0.52 \; . 
\end{equation}  

\noindent 
It is interesting to see here that, due to the heaviness of the
mirror-top quark, the values of the $C$-functions for the top
and mirror-top quarks are comparable in magnitude.

Substituting  the numerical quantities given 
above in Eq.9 constrains  the interference
phase $\omega$ to be  
\begin{equation}
\omega  \approx 57^{o} \pm 96^{o}. 
\end{equation} 

\noindent 
In principle, consistency would require to calculate the quantity
$\rho$ with the NLO $C(m_{i})$ functions. 
It is nevertheless  assumed for simplicity
that the large bottom-quark-scale
uncertainty roughly drops out in the ratio of the mirror-to-standard-model
quantities, and anyway the ignorance of the precise value of $m_{t^{M}}$ 
would render the NLO accuracy superfluous for the purposes of this work. 

It is thus worth noting that there is presently
enough experimental and
theoretical input to mildly constrain some mirror-model parameters. 
One can conclude that potential deviations of similar quantities 
from the standard model predictions can  
be explained within the mirror model, but as stressed before cannot
prove its correctness since there are alternatives ways to get similar
FCNC deviations like supersymmetry or extended technicolor. 
Additional experimental data coming from 
the B-factories will further constrain the mass and mixing parameters
of the mirror model, possibly identifying on the way also novel sources of
CP violation.

\section{Direct tests}

The unique safe  method to prove or falsify the mirror model 
 is obviously to produce
directly the new heavy particles it predicts.  The strongly-interacting
mirror particles introduced in \cite{george1}  are expected to 
either form scalar bound states or propagate freely, since the
mirror-generation group is assumed to break when it becomes strong. 
In any case their masses are 
generally expected to be around  the electroweak scale. 
One should therefore hope that future colliders like the LHC, the NLC and the
muon collider will  
be able to produce these states on-shell and detect them
through their decays.  

\subsection{Vector and fermion fields}

In principle, the existence of 
new vector resonances  should be able to differentiate a
strongly interacting Higgs sector from a perturbative one, like
the one in the standard model or supersymmetry. Decays of
such resonances to a pair $W^{+}W^{-}$ of massive gauge bosons 
would indicate that the new 
fermions carry weak charge and are not $SU(2)_{L}$ singlets as
some see-saw models require \cite{Georgi}. However, the fact that
the strong mirror-generation group is broken, together with the smallness
of the $S$ parameter, 
leads one to suspect that the
theory is not confining even though mirror-fermion condensates are formed, 
and that no vector bound states exist like the rho in QCD or the
technirho in technicolor theories. 

On the contrary, one should expect
elementary massive vector bosons corresponding
to the broken generators of the mirror-generation group. These
would couple strongly to the mirror fermions according to the
mirror-model scenario \cite{george1}, inducing large FCNC between
them. However, it is very questionable whether these vector bosons have
widths which are narrow enough to make them
experimentally detectable. Large widths would dilute
the signal of new decaying vector particles, making their detection at the LHC
conceivable by rather hard.

One should  therefore search for the production of pairs of  
free mirror fermions. Their weak decays  would show  
that these are charged under $SU(2)_{L}$ and not singlets, identifying
with more precision but indirectly the source of possible deviations from the
standard-model values of $g_{R}^{t,b}$ and $|V_{tb}|$. In principle, if mirror
quarks do not pair up with each-other, they 
could  pair-up with ordinary quarks to form 
QCD-singlet scalar bound states, 
so one should hope to see only mirror leptons propagate
freely. Mirror fermions are however expected to decay weakly fast, 
before they have
time to hadronize, in analogy with the top quark.
The expected phenomenology implied by the mirror model at
hadronic colliders is apparently 
so rich that going into a detailed quantitative discussion on
the expected  signals goes beyond the scope of this study. 
A potentially promising process involving zero-spin bound states
is presented in the following. 

\subsection{Scalar fields}

In this subsection, the focus will be on the  
search for  scalar bound states of two mirror fermions.
Parity-violating decays of 
these states    
would indicate that the new fermions are not weak singlets, 
in analogy with
kaon decays which revealed initially the parity-violating
nature of weak interactions more than forty  years ago.  
Furthermore, since the couplings of the scalar bound states are expected
to be similar to  the ones relevant to technipions 
in extended technicolor  theories, a 
most  promising scalar decay would be the one described in \cite{getom}, i.e.
a color-octet and neutral scalar bound state, which we name
``mirror-pion" and denote here by $P_{8}^{M\;0}$, having  a mass equal to
$M_{P}$ and decaying predominantly into a top-antitop quark pair due
to the heaviness of the top quark. 

The mirror pion is mainly produced  via
gluon fusion in very high energy hadronic colliders like the LHC 
due to the large gluon structure functions. 
The relevant decay widths for an $SU(N_{G})$ mirror-generation 
group are 
\begin{equation}
\Gamma(P^{M\;0}_{8} \longrightarrow gg) 
 = \frac{5N_{G}^{2}N_{D}}{384\pi^{3}} \alpha^{2}_{s}(M_{P}^{2})
\frac{M_{P}^{3}}{v^{2}}  
\end{equation}

\noindent  
where $v \approx 250$ GeV
is the electroweak scale, $\alpha_{s}$ the QCD coupling,
$N_{D}=4$ the number of weak mirror doublets,  and
\begin{equation}
\Gamma(P^{M\;0}_{8} \longrightarrow t\bar{t}) \approx
\frac{m^{2}_{t}M_{P}N_{D}}{4\pi v^{2}}\left(1-4\frac{m^{2}_{t}}{M_{P}^{2}}
\right).   
\end{equation}

In the  equation above, $m_{t}$ is the ordinary top-quark mass, and
a CP-conserving effective coupling of the mirror-pion to the top quark
of strength $2m_{t}/v$ has been chosen, in analogy to QCD and
extended technicolor. This should be  a reasonable order-of-magnitude estimate 
for this coupling given the type of mirror-fermion mixing with
the standard-model particles introduced in \cite{george1}. 

The mirror-pion decay into a top-antitop pair dominates over the other
decay modes, making the cross section, which is integrated over an energy bin,
roughly proportional to the
two-gluon decay rate. Since the group $SU(3)_{2G}$
considered here has $N_{G}=3$ instead of
$N_{G}=2$, one would in principle have to multiply the cross-section
results of \cite{getom} by
a factor of $9/4$ for the same mirror-pion and top-quark masses, 
 accelerator c.o.m. energies and experimental cuts.  

Note however that, since in the present case the strong mirror-generation
group is eventually broken, one might have three distinct color-octet 
neutral mirror pions with different masses. Therefore, these do  
not interfere coherently, and they are each  produced by gluon fusion 
with $N_{G}=1$. This reduces the predicted signal for each mirror pion
roughly by a factor of  1/4 compared
to the one in \cite{getom}. Moreover, only the heaviest mirror pion
should be expected to have the  large coupling to the top quark
assumed above due to its mixing with the mirror-top quark, and this mirror
pion should be the main subject of our interest in the following. 

The mirror-pion mass receives the same type of contributions
as in \cite{getom}, with the only difference that 
the extended-technicolor interactions
there have to be replaced by the broken mirror-generation-group interactions
in the present context.
It can be therefore assumed to 
lie in the same mass range as in that reference, i.e. to have a mass around
350-550 GeV.
For the case of masses given by $m_{t}=175$ GeV, $M_{P}=450$ GeV
and c.o.m. energies of 14 TeV which are planned
at the LHC, one should expect an integrated
$t{\bar t}$ production cross-section of about 
\begin{equation}
\sigma_{M}(pp \longrightarrow P^{M\;0}_{8} \longrightarrow t{\bar t})
\approx 10\;\;{\rm pb} 
\end{equation}

\noindent
for the mirror pion for an energy bin of $\pm 10$ GeV around 
its mass and a rapidity cut with $Y=2.5$ \cite{getom}. 

This result should be compared with a QCD $t{\bar t}$
background of around 70 pb's for the same
energy and rapidity cuts.
For the planned LHC luminosities of about
$10^{34} cm^{-2}s^{-1}$, this should guarantee roughly one signal event and
7 background events per 10 seconds  and thus rich statistics. The small
signal-to-background ratio would nonetheless make this enhancement
more difficult to observe.  

In connection with the aforementioned
large deviations of the quantities $|V_{tb}|$ and $g_{R}^{t}$ from their 
standard-model predictions  and with the absence of vector
resonances, such a 
$t-{\bar t}$ production enhancement due to a mirror pion, or other mirror-model
signals at hadronic colliders, 
 could be differentiated in principle from 
 usual technicolor signatures or signals coming from
alternative  dynamical electroweak-symmetry-breaking scenarios. 
 However, it could still be by itself
rather difficult  to prove that the fermions producing these signals 
have indeed mirror- and not standard-model-type 
or singlet weak-charge assignments. 

To acquire this important piece of information, 
one needs to probe 
the weak charges of the new fermions directly, and 
the next subsection is devoted to this quite crucial issue.
To end this subsection on mirror pions, one could also add that  similar 
``higgs"-like scalar resonances would also find a very good and clean
testing ground in the planned muon collider, since the corresponding processes
are proportional to the square of the muon - instead of the electron -
mass $m_{\mu}^{2}$. 

\subsection{The forward-backward asymmetry}

The measurement of the forward-backward asymmetry of mirror fermions
at lepton colliders due to the interference of  lepton-annihilation
 processes into a photon and a $Z^{0}$ boson 
 would be the ultimate proof of the mirror model.
This asymmetry 
has  an opposite sign from the asymmetry of standard-model-type 
fermions expected in 4-generation models for instance. 
However, since the asymmetry also changes sign going from 
fermion masses below the $Z^{0}$ peak to masses above it, mirror fermions
are expected to 
exhibit  a forward-backward asymmetry of the same (negative)
sign as the one observed for the standard-model fermions, except for the
(positive) one corresponding to the top quark that should  manifest itself
at the NLC. 

As already said, 
the strong mirror-generation group is broken, so one can imagine that 
it does not confine, although it gives dynamical masses to the
mirror fermions. Thus, one could expect
to observe an asymmetry of unconfined mirror fermions. 
This is a crucial assumption on which the analysis that
follows is based.  One could also try to 
observe an asymmetry in charged mirror-pion  pair production involving
charged mirror leptons, but the
mirror leptons produced in the fragmentation process isotropically
would, in analogy with light quarks in QCD, most likely wash-out the effect. 

However, since mirror fermions in this model are strongly interacting, 
large radiative effects  make it difficult to observe this asymmetry. 
A $2$-TeV leptonic
collider would already be probing the lower energy  
limits of a possible asymmetry measurement as will be seen shortly, 
since the first-order $\alpha_{G}$
corrections there, where $\alpha_{G}$ is the mirror-generation-group coupling,
are already on the order of $24\%$. Even lower energies would 
allow the mirror-generation-group coupling, along with mirror-fermion
mass-threshold effects,
to more or less smear the directional
information of the two mirror fermions initially produced, possibly 
producing pairs of oppositely charged scalar bound states in
a roughly spherically-symmetric way. 

It would therefore seem beyond any
reasonable hope to measure any mirror-fermion forward-backward asymmetry
at the NLC, unless the mirror-generation group becomes strong
at energies closer to the weak scale than to 1 TeV, something which
is conceivable but difficult, or 
some other - yet-to-be-understood - aspects of the
broken mirror-generation group come into play. This collider could 
concentrate its effort on the also very important task  of
producing these new fermions, which
would at least support the existence of additional generations, if not 
revealing their precise weak-charge assignments.
One should anyway not exclude {\it a priori}
the detection of a small mirror-fermion
forward-backward asymmetry at the NLC. 

Obviously, what is  really
needed is  energies high enough so that the mirror-generation-group 
coupling becomes weak and mirror-fermion mass-threshold effects small. 
Ideally,   a leptonic collider with  
c.o.m. energy $E = \sqrt{s}$ of about 4-10 TeV would be required. 
Such an effort would be reminiscent of
the PEP-PETRA-experiments era of the 80's \cite{PEP}, but with
energies three orders of magnitude higher. 
It should be stressed that the planned muon collider \cite{muon} 
would be a perfect candidate for such a high-energy facility.

Since the up- and down-type 
members of mirror-quark weak doublets have electromagnetic charges 
opposite in sign, the forward-backward 
asymmetry effect would be diluted unless one choses a mirror quark with enough
mass separation from the next lightest and the next heaviest. This 
mass-degeneracy lifting
would reduce heavier mirror-quark contamination and  facilitate flavor
identification in analogy with the bottom- and charm-quark asymmetries in
the standard model. 

To avoid this difficulty, it would be preferable to
measure the forward-backward asymmetry of mirror leptons instead of
mirror quarks. 
In such a
case, only the charged leptons would exhibit an asymmetry if
the mirror neutrinos are Majorana, i.e. if $\nu^{M}={\bar \nu^{M}}$, 
which is the case in \cite{george2}
and  in the following. Mirror leptons could also be
experimentally 
easier accessible since on general grounds they are expected to be
lighter than mirror quarks. 
A conceivable process of considerable interest 
would then be the following: 
\begin{equation}
\mu^{+}\mu^{-} \longrightarrow \gamma, Z^{0\;*} 
\longrightarrow l^{M\;+}l^{M\;-}. 
\end{equation} 

Effects due to the non-relativistic nature of quarks
are quite important in similar standard-model processes, 
where quark current masses can be much larger than
the QCD scale, as is the case for the charm, bottom and top quarks.
The mirror fermions considered 
here  however have purely dynamical masses, and by the
time one reaches energies where mirror-generation-group
strong-coupling effects can
be controlled perturbatively, the mirror fermions are to a
very good approximation relativistic.  

\begin{figure}[p]
\vspace{4.1in}
\includegraphics{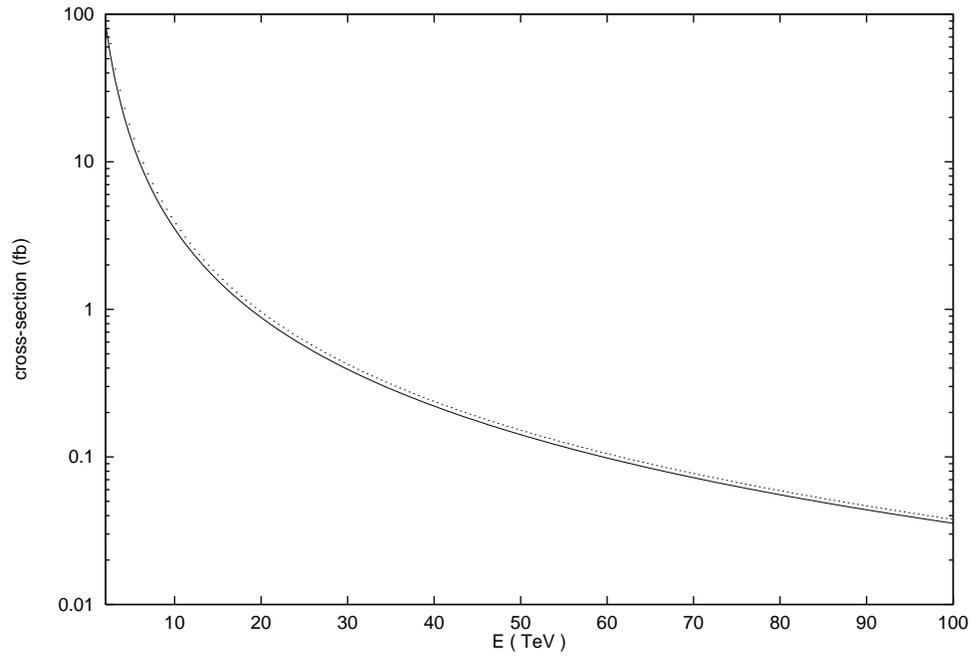}
\caption{The scattering cross-section for the process
$\mu^{+}\mu^{-} \longrightarrow l^{M\;+}l^{M\;-}$
as a function of the c.o.m. energy E.
The dotted line shows the result     
after taking the strong mirror-generation-group effects into account.} 
\end{figure}

\begin{figure}[p]
\vspace{4.1in}
\includegraphics{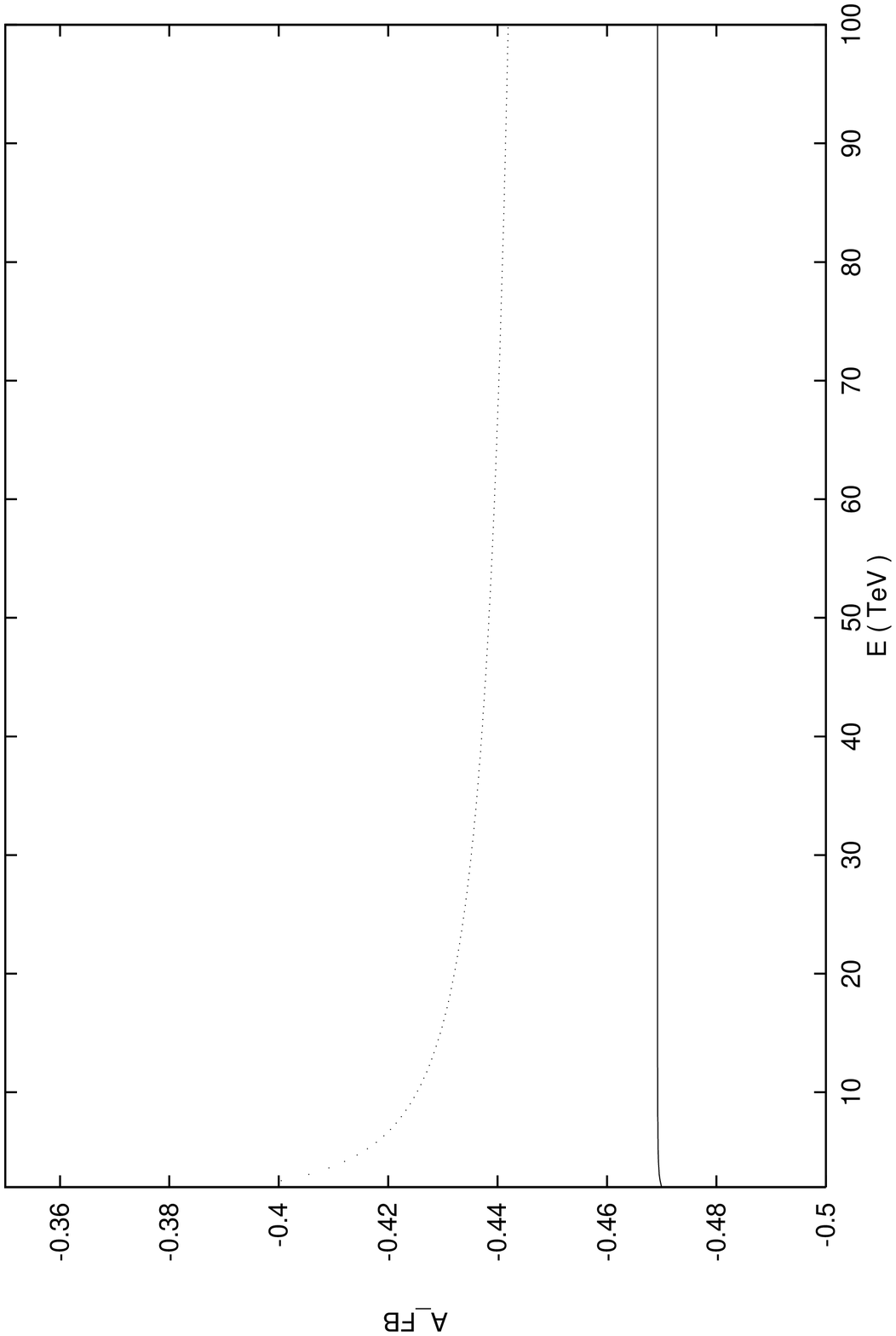}
\caption{The forward-backward asymmetry of the mirror leptons. 
The dotted line shows the
decreased asymmetry due to the strong 
mirror-generation-group effects. The dependence
of this line with energy is due to the logarithmic running of the 
corresponding gauge coupling.} 
\end{figure}

Therefore,   the velocity of the mirror fermions of mass 
$m \approx 200-600$ GeV is taken in the following to be
$\beta = \sqrt{1-m^{2}/s} \approx 1$. 
In this case, the production cross-section and the 
forward-backward asymmetry for the generic process 
$l^{+}l^{-} \longrightarrow f^{+}f^{-}$ via a photon and a $Z^{0}$ gauge
boson for Dirac leptons $f^{\pm}$ in the final state  
are given respectively by the following relations:  
\begin{eqnarray}
\sigma(l^{+}l^{-} \longrightarrow f^{+}f^{-}) 
& = & \frac{4\pi N_{G}\alpha^{2}}{3s}Q_{1} \nonumber \\  
A^{0}_{FB} &=& \frac{3Q_{3}}{4Q_{1}} , 
\end{eqnarray}  

\noindent 
where $N_{G}$ is the number of ``colors" carried by the final-state 
leptons, and the  quantities $Q_{1,3}$ are defined as \cite{aguila} 
\begin{eqnarray}
Q_{1} & = & \frac{1}{4}(|Q_{LL}|^{2}+|Q_{RR}|^{2}
+|Q_{LR}|^{2}+|Q_{RL}|^{2}) \nonumber \\  
Q_{3} & = & \frac{1}{4}(|Q_{LL}|^{2}+|Q_{RR}|^{2}
-|Q_{LR}|^{2}-|Q_{RL}|^{2}).     
\end{eqnarray} 

At these high energies the mirror-generation group can
be considered to be unbroken, since it is in principle impossible to
distinguish between the mirror leptons $e^{M}, \mu^{M}$ and $\tau^{M}$, 
so one has  $N_{G}=3$.  
The quantities $Q_{JK}$ introduced above are equal to    
\begin{equation}
Q_{JK} = q^{l}_{J}q^{f}_{K} + 
\frac{Q^{l}_{J}Q^{f}_{K}}{s_{W}^{2}c_{W}^{2}}
\frac{s}{s-m_{Z}^{2}+im_{Z}\Gamma_{Z}}, 
\end{equation}

\noindent with 
$s^{2}_{W}=1-c_{W}^{2}=sin^{2}{\theta_{W}}$
the Weinberg angle, $m_{Z} \approx 91$ GeV and $\Gamma_{Z}\approx 2.5$ GeV
the $Z^{0}$ mass and width
respectively, $q^{l,f}$ the electromagnetic charges of the initial and 
final leptons, and 
\begin{equation}
Q^{l,f}_{J}=I_{3\;J}^{l,f} -s^{2}_{W}q_{J}^{l,f}
\end{equation}

\noindent 
the electroweak charges of the initial and final leptons 
of isospin $I_{3\;J}^{l,f}$ respectively, with
$J, K =L,R$. 

The energies considered
here are so high that the $Z^{0}$-boson mass and decay width could be safely 
neglected from the start. For the same reason, 
contributions to this process from scalar resonances like effective
``higgses" can also be neglected since, for the muon collider for 
instance, one has $\left(\frac{m_{\mu}}{v}\right)^{2} \ll \alpha^{2}$.
It should be further reminded that, in the formulas above, $\alpha$ and
and $s^{2}_{W}$ are energy-dependent due to the renormalization-group
equations, which in the present framework 
take also mirror fermions  into account \cite{george2}. 

Taking the mirror-model quantum-number assignments into consideration, 
the lepton charges are given by  
\begin{eqnarray}
&&Q^{l}_{L} = Q^{f}_{R} = -\frac{1}{2} + s^{2}_{W}, \;\;\;\;\;\;\;\;\;\;\;\;\;
\;\;\;\;\;\;\;\;\;\;\;\;\;\;\;\;\;\;\;\;\;
Q^{l}_{R} = 
Q^{f}_{L} = s^{2}_{W} \nonumber \\ 
&&q^{l}_{L} = q^{l}_{R} =  q^{f}_{L} =  q^{f}_{R} = - 1. 
\end{eqnarray}

\noindent  
The smallness of the  mixing of the mirror leptons with their standard-model
counterparts  allows one to safely
neglect in this calculation the fact that mirror leptons are not
pure mass eigenstates.

Moreover, it is easy to check that
having final fermions with mirror charge assignments \cite{george1}
instead of standard-model type ones corresponds to a
$L \leftrightarrow R$ interchange in the second subscript of the
quantities $Q_{JK}$ with $J, K = L, R$,  and
a resulting change in the overall sign of $A^{0}_{FB}$. The sign of the
forward-backward asymmetry of the final-state fermions
provides therefore a test which is unique in its importance for the
experimental verification of the mirror model. 

Including first-order corrections due to the strong mirror-generation-group 
coupling $\alpha_{G}$ and neglecting possible effects due to
the fact that the mirror-generation group is eventually broken, one has  
a production cross-section enhanced by a factor of 
$\left(1 + \frac{\alpha_{G}(s)}{\pi}\right)$, and a corresponding 
reduced asymmetry result equal to  
\begin{equation}
A_{FB} \approx A_{FB}^{0}
\left(1 + \frac{\alpha_{G}(s)}{\pi}\right)^{-1},
\end{equation}

\noindent where the running of the coupling of the mirror-generation group  
$SU(3)_{2G}$ is taken to be described by the relation  
\begin{equation}
\alpha_{G}(s)=\frac{1}{1+\frac{17 \ln{(s/{\rm TeV^{2}})}}{12\pi}} . 
\end{equation}

Even though the first-order
correction is substantial, the precise magnitude of the
second-order correction is usually debated but
quite smaller \cite{9905424}, and can anyway not answer by itself the question 
on whether the full $SU(3)_{2G}$-corrected 
result is much smaller or larger than the
first-order result.   At energies of 
 4 TeV one finds $\alpha_{G} \approx 0.44$, which is still a pretty large
parameter  to do perturbation theory with. 
The magnitude of the first-order result at energies 
close to 1 TeV makes
clear that the numerical values presented in the following, especially for the 
lower energies, should be seen more as
qualitative estimates  than as precise predictions.

The results for the cross-section and the forward-backward asymmetry
as functions of c.o.m. energy with and without first-order
mirror-generation-group corrections
are shown in Figures 1 and 2 respectively. 
Energies from 2  to  100 TeV are considered. 
Such high energies have already been discussed in the context of
the muon collider, even though they are clearly referring to 
experiments in the far future. 
Figure 2 makes clear why the NLC
energies would make such a measurement very difficult because of the
large $SU(3)_{2G}$ corrections, and why the muon
collider would be a very good solution.

Unlike the case of PEP-PETRA experiments, 
$A_{FB}^{0}$ is not damped by the $Z^{0}$-boson mass and
it reaches  an almost constant value. At around 10 TeV, one finds  
$A^{0}_{FB} \approx - 0.47$. 
At such very high energies, the forward-backward  asymmetry
has still a very mild energy dependence due to the
renormalization of $\sin^{2}\theta_{W}$. 
The $SU(3)_{2G}$-corrected
result has a slightly  stronger energy dependence due to the
mirror-generation-group coupling renormalization.
For energies on the order of  10 TeV, one finds 
\begin{equation}
A_{FB} \approx -0.42, 
\end{equation}

\noindent
i.e. the forward-backward asymmetry is reduced by roughly $11\%$. 

Even though for larger  
energies the $SU(3)_{2G}$ corrections would very slowly become smaller, 
the cross-section decreases fast, leaving the phenomenological 
analysis with less statistics.
With the planned muon-collider  luminosities of around 
$10^{34}cm^{-2}s^{-1}$ \cite{muon}, 
a 4-TeV machine would produce about $10^{3}$ events per year, while
a hypothetical 100-TeV machine would need luminosities
roughly 3 orders of magnitude larger in order to have comparable statistics. 

By using mirror leptons, 
one avoids any QCD corrections to the theoretical predictions.
Moreover, since the energies 
of interest here are much higher than 
the $Z^{0}$-peak \cite{9707}, 
electroweak corrections are    
ignored in this first approach to such processes.
For the same reason, initial-lepton polarization is not essential for 
this measurement. 
The QED corrections in their turn 
could be substantially reduced by an angular acceptance cut
like $|\cos{\theta}|< \Theta $, where $\theta$ is the scattering angle
between the initial and final fermions. Taking  $\Theta =0.8$ as is frequently
done would however reduce (independently of the QED corrections) 
the cross-section by $1 - \frac{3\Theta +\Theta^{3}}{4} \approx 27\%$ and the
forward-backward asymmetry values by
 $1 - \frac{4\Theta^{2}}{3\Theta+\Theta^{3}} \approx 12\%$. 

Such a cut would also reduce the forwardly-peaked $W^{+}W^{-}$
standard-model background, since the $W$ bosons 
coming from mirror-lepton weak decay are
isotropically distributed. This is the most important background
process, since the mirror leptons, assumed here to be heavier than the
mirror neutrinos, decay predominantly to a mirror neutrino and a W boson. 
The mirror neutrinos are expected to decay in their turn via the weak  
process $\nu^{M}_{j} \longrightarrow W^{\pm} l^{\mp}$, 
with a decay rate roughly equal to 
\begin{equation}
\Gamma \approx \frac{G_{F}m^{3}_{\nu^{M}_{j}}}{8\sqrt{2}\pi}
|U_{lj}|^{2} 
\end{equation}

\noindent where  the index $j$ runs over the mirror neutrinos and
$U_{lj}$ is an element of 
the lepton CKM-type mixing matrix discussed in \cite{george2}. 

The standard-model background 
could be further reduced by 
choosing pairs of W-bosons which are not co-linear and co-planar, since
part of the momentum of the process
is carried by the mirror neutrinos. This should be
done in conjunction 
with a  hard-photon cut, to ensure that such events do not
come from higher-order
standard-model interactions. Further background reduction can be
achieved by a visible-energy cut, since there is missing energy from the
jets carried by the mirror neutrinos. All these cuts   should produce
a signal-to-background ratio large enough  to facilitate the analysis which
would verify or falsify the mirror model.

\section{Conclusions}

The phenomenological ``excursion" of
this work, by no means claiming to have spanned the whole spectrum of
conceivable tests,
has shown  that several experimental  
consequences of the strongly-interacting
mirror-fermion model can be tested in various present and future high-energy
facilities. Indications for the existence of mirror fermions could 
already be coming from the values of $\delta g^{b}_{R}$ and the 
electroweak precision 
$S$ and $T$ parameters. Future measurements of $|V_{tb}|$ and 
$\delta g^{t}_{R}$ at the Tevatron III and the NLC could provide further
indirect evidence for the
model. Large deviations of quantities related to 
$b \longrightarrow s \gamma$ or other
B-meson processes from their standard-model predictions could also 
be made consistent with or constrain
such an extension of the standard model.  

Direct production of new fermions and scalar bound states at
hadronic colliders would
offer more substantial evidence for a new strongly-interacting
sector, but it would be hard to distinguish unambiguously
the resulting signals
from other theories like technicolor. 
The ultimate confirmation of the mirror model  would come
from a future large linear collider 
with c.o.m. energies  around 4-10 TeV, in much the same way
that the LEP/SLC experiments confirmed the standard model in the
90's. Such a high-energy collider would be able to probe the precise 
chiral structure of the new fermions via
their forward-backward asymmetry $A_{FB}$, assuming of course that they
can propagate freely and are not always confined in bound states. 

In the latter case one would have to resort to other methods for 
determining the chirality of the new fermions, not different in
spirit perhaps from the methods used to determine the chirality of
the weakly-charged electrons and protons before the forward-backward
asymmetry of standard-model fermions was measured. 
In the former case however, the muon collider would be a very good
candidate for such a high-energy facility, so it would be very
encouraging if
it  finally proved to be technologically feasible. The discussion 
presented implies anyway  
that the processes studied in this work deserve further detailed study
within the context of specific particle-physics experiments. 

\noindent {\bf Acknowledgements} \\
I thank P. Gambino, N. Maekawa and L. Silvestrini for useful discussions. 
This research is supported by an {\it Alexander von Humboldt} Fellowship.

\newpage
\noindent Figure Captions

\noindent Figure 1:

\noindent The scattering cross-section for the process
$\mu^{+}\mu^{-} \longrightarrow l^{M\;+}l^{M\;-}$
as a function of the c.o.m. energy E.
The dotted line shows the result     
after taking the strong mirror-generation-group effects into account. 

\noindent Figure 2:

\noindent The forward-backward asymmetry of the mirror leptons. 
The dotted line shows the
decreased asymmetry due to the strong 
mirror-generation-group effects. The dependence
of this line with energy is due to the logarithmic running of the 
corresponding gauge coupling. 


\begin{thebibliography}{99}
\bibitem{george1} G. Triantaphyllou, Technical University Munich 
Preprint No. TUM-HEP-326/98, September 1998, hep-ph/9811250. 
\bibitem{george2} G. Triantaphyllou, 
to appear in {\it Eur. Phys. Jour.} {\bf C}, hep-ph/9901346.
\bibitem{Barr} S.M. Barr, D. Chang and G. Senjanovic,
{\it Phys. Rev. Lett.} {\bf 67}, 2765 (1991). 
\bibitem{Field} J.H. Field, {\it Phys. Rev.} {\bf D 58}, 093010 (1998);
Universit\'{e} de Gen\`{e}ve
Preprint No. UGVA-DPNC 1998/09-179, September 1998,
hep-ph/9809292; {\it ibid} October 1998, hep-ph/9810288;
A.K. Grant and T. Takeuchi, UCLA Preprint No.
UCLA/98/TEP/20, July 1998, hep-ph/9807413;
P.B. Renton, Oxford University Preprint No. OUNP-98-08, November 1998,
hep-ph/9811415.
\bibitem{Hil} C.T. Hill {\it et al.}, {\it Phys. Rev.} {\bf D 47}, 
2940 (1993).  
\bibitem{yale} T. Appelquist and G. Triantaphyllou, {\it Phys. Lett.}
{\bf B 278}, 345 (1992).  
\bibitem{Chivu} R. Frey {\it et al.}, FERMILAB-CONF-97-085, April 1997,
hep-ph/9704243. 
\bibitem{Roos} For a review, see J. Maalampi and M. Roos, {\it Phys. Rep.} 
{\bf 186}, 53 (1990). 
\bibitem{Cho} M. Ciuchini {\it et al.}, {\it Phys. Lett.} {\bf B 316},
 127 (1993); {\it ibid}, {\it Nucl. Phys.} {\bf B 421}, 41 (1994); 
P. Cho and B. Grinstein, {\it Nucl. Phys.}
{\bf B 365}, 279 (1991); erratum {\it ibid.} {\bf 427}, 697 (1994);
G. Buchalla, A.J. Buras and M.E. Lautenbacher, {\it Rev. Mod. Phys.}
{\bf 68}, 1125 (1996).
\bibitem{Buras} For a recent review, see
A.J. Buras, Munich Technical University Preprint
TUM-HEP-316/98, June 1998,  hep-ph/9806471.
\bibitem{Fuji} K. Fujikawa and A. Yamada, {\it Phys. Rev.} {\bf D 49},
5890 (1994).
\bibitem{cleo} T. Skwarnicki, talk given at ICHEP98, July 1998, Vancouver. 
\bibitem{Georgi} R.S. Chivukula {\it et al.}, {\it Phys. Rev.} {\bf D 59},
075003, 1999.
\bibitem{getom} T. Appelquist and G. Triantaphyllou, 
{\it Phys. Rev. Lett.} {\bf 69}, 2750 (1992). 
\bibitem{PEP} We give indicatively the references
P. Baringer {\it et al.}, {\it Phys. Lett.} {\bf B 206}, 
551 (88); F. Ould-Saada {\it et al.}, {\it Zeit. Phys.}{\bf C 44}, 567 (89); 
W. Braunschweig {\it et al.}, {\it Zeit. Phys.}{\bf C 48}, 433 (90). 
\bibitem{muon} C.M. Ankenbrandt {\it et al.}, Fermilab
Preprint No. FERMILAB-Pub-98/179, January 1999.
\bibitem{aguila} See for instance F. del Aguila {\it et al.}, 
{\it Nucl. Phys.} {\bf B 297}, 1 (1988). 
\bibitem{9905424} S. Catani and M.H. Seymour, CERN Preprint CERN-TH/99-132,
May 1999, hep-ph/9905424.
\bibitem{9707} U. Baur, S. Keller and W.K. Sakumoto,
{\it Phys. Rev.} {\bf D 57}, 199 (1998).
\end{thebibliography}
\end{document}